Series IV

# A “1+1” Algorithm for the Hamilton Cycle Problem

Heping Jiang[0000-0001-5589-808X]

jjhpjhp@gmail.com

## Abstract

Deciding if a graph is a Hamilton graph, also named the Hamilton cycle problem, is important for discrete mathematics and computer science. Due to no characterization to identify Hamilton graphs effectively, there are no tractable algorithms to solve the Hamilton cycle problem. Grinberg Theorem is a necessary condition only for planar Hamilton graphs. In this paper, based on new studies on the Grinberg Theorem, in which we provided new properties of Hamilton graphs with respect to the cycle bases and improved the Grinberg Theorem to derive an efficient condition for Hamilton graphs, we present a new precise algorithm for deciding Hamilton graphs, named the “1+1” algorithm. Theoretically, the “1+1” algorithm terminates in $\mathcal{O}(|E(G)|^3)$ worst time complexity, where $|E(G)|$ is the size of the given graph $G$.

## 1 Introduction

A Hamilton cycle in a graph is a cycle that contains all the vertices once and a graph contains a Hamilton cycle is called a Hamilton graph. We sometimes use ‘a graph is Hamiltonian’ to express a Hamilton graph. How to identify a Hamilton graph, also called the Hamilton cycle problem (HCP in abbreviation), is a well-known problem in discrete mathematics and computer science.

Despite extensive studies on the Hamilton cycle problem, different from an Euler graph and its algorithm, no 'good characterization' [1] of Hamilton graphs has been found to be an effective condition [2]; therefore, it almost couldn't evade the computational complexity increasing exponentially in the order-of-magnitude at present. In computational complexity, the Hamilton cycle problem is NP-complete [3].

Generally, algorithms for HCP are classified into the precise and the approximate algorithms. The precise algorithms usually mean the back-tracking approaches, by which given some constraints locally we can assess every potential solution in a branch of the searching tree, and eventually find a Hamilton cycle in a graph. Although a back-tracking algorithm can guarantee to find a Hamilton cycle, it clearly takes exponential time in the worst case. On the other hand, approximate algorithms use the heuristic approaches, which perform all the searching by general rules in polynomial time. It possibly gives a wrong solution despite quick executing. Basil Vandegriend [4] and Andrew Chalaturnyk [5] detailed various algorithms for HCP in their theses. In summary, as of now most studies of the algorithms for HCP have been focusing on improving the techniques of algorithms themselves, while there are few the explorations for algorithms based upon new Hamiltonian properties.

In this paper, based upon the studies on new characteristics of Hamilton graphs in [6-8], we concentrate on applying new Hamiltonian properties to the algorithm developments, and propose a new precise algorithm for the Hamilton cycle problem on solvable norm graphs (defined in later section), named the "1+1" algorithm. The name of the algorithm comes from the visual representation of the essential steps to construct a set of cycles (denoted by sub-$\mathcal{G}$) in the given graph. We show that the "1+1" algorithm gives out the decision in $\mathcal{O}(|E(G)|^3)$ worst case time complexity, where $|E(G)|$ is the size of the given graph.

In outline, the "1+1" algorithm begins from selecting an arbitrary cycle in the solution set [7] of a graph, and adds the cycles one by one in the way of 2-common ($v$, $e$) combination [6]; and then, we obtain a set of solution cycles (denoted by sub-$\mathscr{g}$). Afterwards, using a serious methods, showed in the following section, we compare the sub-$\mathscr{g}$, also call the working set in the paper, to the solution of the given graph [6] for obtaining a subgraph $\mathscr{g}$ (a cycle set without removable cycles) of the given graph [8], which we can determine whether the given graph is Hamiltonian or not according to Theorem 1.1 in [8]. Nevertheless, in the procedure of the algorithm, due to we continue to sift through all the possibilities of the working sets with the result that the given graph is Hamiltonian or not, the algorithm eventually terminates after ergodic searching and comparing in polynomial time. We will detail the algorithm in the following sections.

In section 2, we provide the definitions and notions mentioned in this paper. Section 3 includes all the proofs of propositions and lemmas in the methods applying in the "1+1" algorithm. In section 4, we give the theorem of the "1+1" algorithm and its proof, which includes the steps of the algorithm.

## 2 Definitions and notions

Since this paper is a part of a collection, we hope readers refer to the papers [6-8] for some notions, definitions and notations, and results not mentioned here. For the terminologies not mentioned in this paper, please refer to [11-14].

All numbers throughout the paper is positive integer, denote by $\mathbb{Z}^+$.

In [6] and [7], we defined the solvable graphs and normal graphs, respectively. In this paper, $G$ denotes a solvable normal graph. $B(G)$ denotes the minimal cycle basis of $G$, and $|B(G)|$ the number of cycles in $B(G)$. $S_x$ is called a particular solution of $G$ such that $S_x = \{C_i \mid 3 \le i \le |V(G)| \text{ and } \sum(i|C_i|-2|C_i|)=|V(G|)-2\}$[1], where $x \in \mathbb{Z}^+$. $\boldsymbol{S}$ is called a general solution of $G$ such that $\boldsymbol{S} = \{S_x \mid x \in \mathbb{Z}^+\}$. A particular solution cycle, denoted by $p$-$\boldsymbol{C}$, is a cycle in $S_x$. A general solution cycle, denoted by $g$-$\boldsymbol{C}$, is a cycle in $\boldsymbol{S}$.

[1] $\sum(i|C_i|-2|C_i|)=|V(G|)-2$ is the equation of a graph in abbreviation.

In [6], we define the 2-common ($v$, $e$) cycle set as a cycle combination such that every pair of joint cycles, *e.g.*, cycle $A$ and $B$, satisfying $|V(A)\cap V(B)|$ = 2 and $|E(A)\cap E(B)|$ = 2; and the 2-common ($v$, $0$) cycle set denotes a cycle combination such that every pair of joint cycles, *e.g.*, cycle $A$ and $B$, satisfying $|V(A)\cap V(B)|$ = 2 and $|E(A)\cap E(B)|$ = 0. Sometimes we also call them a 2-common ($v$, $e$) combination and a 2-common ($v$, $0$) combination.

$\mathscr{g}$ is a subgraph obtained by deleting all removable cycles from a basis of $G$. Let $sub$-$\mathscr{g}$ be a cycle set of $G$ produced by adding the cycles $C_i$, where $C_i \in S_x$, $3 \leq i \leq |V(G)|$, one by one in a way of 2-common ($e$, $v$) combination. We also call $sub$-$\mathscr{g}$ a working set in this paper.

We use a match for a comparison of two (cycle) sets, and write $A = B$ as the set $A$ matches with the set $B$. Particularly, in this paper, we use a match for comparing a $sub$-$\mathscr{g}$ to $S_x$ with order, size and number of cycles.

$\alpha$ denotes an isolated vertex not covered by a set $sub$-$\mathscr{g}$, and $C$ denotes a cycle covering the vertex $\alpha$. For a $sub$-$\mathscr{g}$+$C$, we say a solution cycle is a missing cycle if it is a $p$-$\boldsymbol{C}$ but not in the $sub$-$\mathscr{g}$+$C$, and a solution cycle is a redundant cycle if it is not a $p$-$\boldsymbol{C}$ but in the $sub$-$\mathscr{g}$+$C$. $\Delta C$ denotes an unassigned cycle. We write $sub$-$\mathscr{g}$+$C$+$\Delta C$ if $\Delta C$ is a missing cycle, and write $sub$-$\mathscr{g}$+$C$−$\Delta C$ if $\Delta C$ is a redundant cycle. $E_\beta$ is an edge shared by $\Delta C$ and $C$ such that $\Delta C$+$C$ is a 2-common ($v$, $e$) combination, where $C$ is called a basal cycle and $C \in sub$-$\mathscr{g}$+$C$. $\beta$ denotes a vertex on the $\Delta C$ not incident to $E_\beta$. $\beta_2$ denotes a boundary $\beta$ of degree 2. Let $C_\Delta$ be a cycle on $\beta$ such that $V(G–C) = V(G)$ for $G – C$ and $C_\Delta \neq \Delta C$. We write $|\beta_2| = 0$ as that the vertex $\beta$ cannot be reduced to $\beta_2$ by deleting $C_\Delta$. We use the $\beta_2$-construction for an operation such that the vertex $\beta$ can be reduced to $\beta_2$ by deleting all cycles not including the boundary vertex of degree 2 in the graph G. Let $\bowtie$ denote a cut vertex.

**Lemma 1** *Let G be a solvable norm graph. Let* $\bowtie$ *be a cut vertex.*

*Then, for cycles C and C' in G, there is a vertex* $\bowtie$ *connected C and C' if* $|V(C)| + |V(C')| = | V(C) \cup V(C') | + 1$.

*Proof.* Let $|V(C)|$ denote the order of cycle $C$, $|V(C')|$ the order of cycle $C'$. If *C and C'* connect with a common vertex, then the order of their combination is $|V(C) \cup V(C')|$. We clearly have $|V(C) \cup V(C')| < |V(C)| + |V(C')|$. By the definition, the common vertex is a cut connected *C and C'*. Thus, the statement holds. ■

## 3 Methods

By Lemma 1.1 and 1.2 in [6], unsolvable graphs are non-Hamiltonian, thus, for deciding the graphs' Hamiltoncity, we only consider the solvable graphs; by the definition of the reduced graph [8], for simplification of the Hamilton graphs decision, we consider the norm graphs. Thus, we here only discuss the solvable norm graphs.

By Lemma 1.2 [6], we showed that there are two types of cycle sets in a solvable graph,

(a) A 2-common ($v$, $e$) cycle set,

(b) A 2-common ($v$, $0$) cycle set.

In [7], with investigating the 2-common ($v$, $0$) cycle set, we characterized the figure of the cycle $C_k$ in it, and showed that a solvable norm graph is Hamilton graph, if and only if, $|C_k| = 0$.

In [8] we proved that a norm graph is non-Hamiltonian, if and only if, $\mathcal{G} \cong K_{2,3}$, where $\mathcal{G} \cong K_{2,3}$ denotes $\mathcal{G} \approx^{2} K_{2,3}$ such that only one subgraph $K_{2,3}$ induced from $\mathcal{G}$.

Accordingly, motivated by works of [6], [7], and [8], for solvable norm graphs, we suggest to construct a cycle set that matches with the solution set $S_x$ of the given graph, so that the cycle set is either a 2-common ($v$, $e$) cycle set or a 2-common ($v$,$0$) cycle set; and hence, by identifying whether or not there exists a subgraph $K_{2,3}$ in it, we can eventually determine the Hamiltoncity of the given graph.

---

[2] The notation "≈" means "to be homeomorphic to" [8].

The main methods applied in our algorithm include as followings. First, beginning from any selected cycle in a given solution set $S_x$, we continue to add the other cycles one by one in the way of 2-common ($e$, $v$) combination; the result of the adding is a set of solution cycles *sub-*$\mathcal{G}$. Second, if the *sub-*$\mathcal{G}$ itself is not a Hamilton cycle set, then there must exist isolated vertices $\alpha$; and we select cycles $C$ in the graph to cover them so that the *sub-*$\mathcal{G}$+$C$ covers all vertices in the given graph. Third, if the *sub-*$\mathcal{G}$+$C$ itself is not a Hamilton cycle set, then we use the unassigned cycles $\Delta C$ to match the *sub-*$\mathcal{G}$+$C$ with the solution set $S_x$; and we have a result such that either the *sub-*$\mathcal{G}$+$C$ with $|\Delta C|=0$ or the *sub-*$\mathcal{G}$+$C$ with $|\Delta C|\neq 0$. For $|\Delta C|=0$, it means the *sub-*$\mathcal{G}$+$C$ itself is a non-Hamilton cycle set (because there have inside vertices in the *sub-*$\mathcal{G}$+$C$). For $|\Delta C|\neq 0$, we go to the forth step, to certain whether or not a selected $\Delta C$ is a missing or a redundant cycle in producing the *sub-*$\mathcal{G}$+$C$; and then we obtain the *sub-*$\mathcal{G}$+$C \pm \Delta C$. Finally, based on a characterization with respect to subgraph $K_{2,3}$, we relate the Hamiltoncity of the given graph to the existence of the isolated vertex $\alpha$ in the $\beta_2$-construction for all the cycles $\Delta C$.

Clearly, the main procedure of the algorithm begins from the working set *sub-*$\mathcal{G}$. So, for simplification, we assume that the set *sub-*$\mathcal{G}$ is given in section 3.

### 3.1 Adding $C$

According to the definition of the *sub-*$\mathcal{G}$, we have

**Proposition 1** *Let G be a solvable norm graph. If* $|C| = 0$ *for the sub-*$\mathcal{G}$*, then G is a Hamilton graph.*

*Proof.* Let $G$ be a solvable norm graph. By the definition, a *sub-*$\mathcal{G}$ is a 2-common ($v$, $e$) combination. While $|C| = 0$ implies that the *sub-*$\mathcal{G}$, 2-common ($v$, $e$) cycle set, covers all vertices of $G$. Thus, the sum[3] of the whole cycles is a Hamilton cycle in $G$. ■

[3] We denote by A + B = (A ∪ B) \ (A ∩ B) the *sum* of subgraphs A and B in a graph. [6]

For $|C| \neq 0$, it means there exists the vertex $\alpha$ not covered by the *sub-g* in $G$. Note that all the cycles belong to a basis of $G$, then, there must have a cycle $C$ in $G$ covering the vertex $\alpha$. And then, we derive the *sub-g* $+ C$. The selected cycle $C$ shows the following property.

**Proposition 2** *Let G be a solvable norm graph. Let* $\alpha$ *be an isolated vertex not covered by the sub-g, where sub-g* $\subseteq G$. *Let* $\boldsymbol{S}$ *be a solution set of G such that* $\boldsymbol{S} = \{S_x \mid x \in \mathbb{Z}^+\}$. $S_x = \{ f_i \mid 3 \leq i \leq |V(G)|$ *and* $\sum(i|f_i| - 2|f_i|) = |V(G|) - 2\}$. *Let* $C$ *be a cycle covering the vertex* $\alpha$ *in G. Then,* $C$ *is a g-*$\boldsymbol{C}$ *if* $|C| \neq 0$.

*Proof.* Let $C$ be a cycle covering the vertex $\alpha$ in a solvable norm graph $G$. Assume $C$ is not a *g*-$\boldsymbol{C}$. Then, the cycle $C$ is neither in the set *sub-g* nor in the set $S_x$. Then, cycle $C$ is definitely in the co-solution set of $G$, which is a cycle to be deleted. That means the equation of $G$ does not include the vertex $\alpha$, which contradicts to that $G$ is a solvable graph. Hence, the statement holds. ■

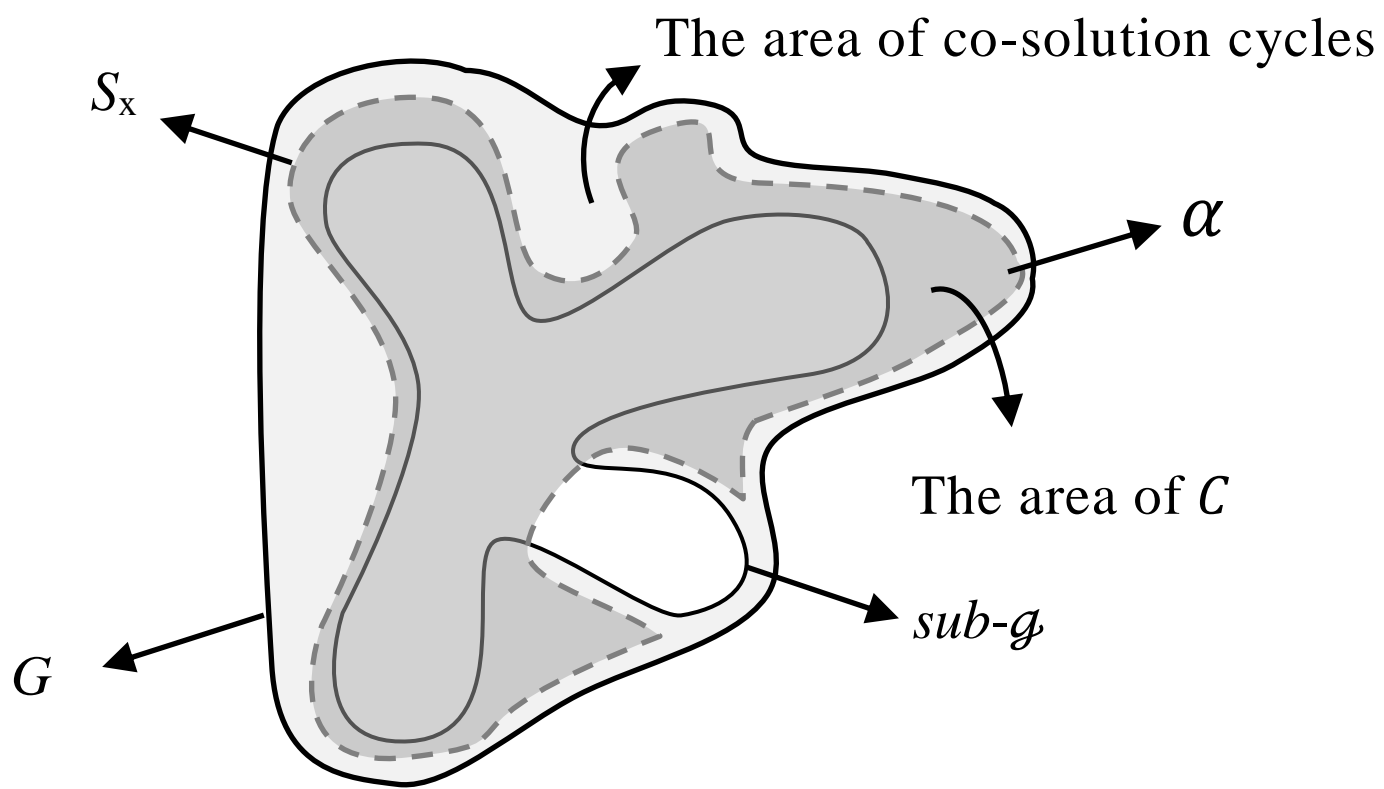

Figure 1. An overlapping diagram of a graph $G$, a solution $S_x$ and the set *sub-g*. The light grey area around by bold line represents a graph $G$, the dark grey area around by dotted line is a solution area $S_x$, and the white area around by light line denotes the set *sub-g*.

## 3.2 Adding $\pm\Delta C$

By the definition, the process to select cycles for generating a cycle set *sub-g* is stochastic, so it is uncertain for the *sub-g*$+C$ to match with $S_x$ for $|C| \neq 0$. Therefore, by the method of adding $\pm\Delta C$, we adjust the cycle structure in the *sub-g*$+C\pm\Delta C$ for matching to $S_x$.

However, the following claim holds if $|\Delta C| = 0$.

**Lemma 2** *Let G be a norm solvable graph. Let $S_x$ be the cycles in the solution sets of G such that $S_x = \{f_i \mid 3 \leq i \leq |V(G)|$ and $\sum(i|f_i| - 2|f_i|) = |V(G|) - 2\}$. Let $C$ be a cycle covering the vertex $\alpha$ in a solvable norm graph G. Then, sub-$\mathscr{g}$ + $C = S_x$, if $|\Delta C| = 0$.*

*Proof.* Suppose we obtain a cycle set *sub-*$\mathscr{g}$+$C$ such that $|\Delta C|=0$. By the definition, the cycle $\Delta C$ is a missing cycle or a redundant cycle to the set of *sub-*$\mathscr{g}$+$C$. Thus, for a norm solvable graph, $|\Delta C|=0$ implies the *sub-*$\mathscr{g}$+$C$ covers all vertices of $G$. Hence, the *sub-*$\mathscr{g}$+$C$ is a solution set of $G$, that is *sub-*$\mathscr{g}$+$C = S_x$. ■

From Lemma 1, we immediately derive the following corollary,

**Corollary 1** *Let G be a norm solvable graph. Then, G is a non-Hamilton graph if $|\Delta C| = 0$.*

*Proof.* By Lemma 2, if $|\Delta C|=0$, then we have *sub-*$\mathscr{g}$ + $C = S_x$, where $|C| \neq 0$. Note that there must exist some inside vertices generated from adding the cycle $C$, which means the sum of the *sub-*$\mathscr{g}$+$C$ does not cover these inside vertices, and the set is not Hamiltonian. In addition, note that the *sub-*$\mathscr{g}$+$C$ is a 2-common $(v, e)$ cycle set of $G$ either or a 2-common $(v, 0)$ cycle set of $G$. Thus, the *sub-*$\mathscr{g}$+$C$ must be a 2-common $(v, 0)$ cycle set of $G$. Hence, G is non-Hamilton graph. ■

For $|\Delta C| \neq 0$ and *sub-*$\mathscr{g}$ $+C = S_x$ where $|C| \neq 0$, we have the following lemma.

**Lemma 3** *Let G be a norm solvable graph. Let $S_x$ be a solution of G such that $S_x = \{f_i \mid 3 \leq i \leq |V(G)|$ and $\sum(i|f_i| - 2|f_i|) = |V(G)| - 2\}$. $\Delta C$ denotes an unassigned cycle. if $|\Delta C| \neq 0$, then sub-$\mathscr{g}$ + $C \pm \Delta C = S_x$, where $|C| \neq 0$.*

*Proof.* By the definitions, if $|C| \neq 0$, then the *sub-*$\mathscr{g}$ + $C$ covers all the vertices of $G$; and if $|\Delta C| \neq 0$, then we obtain a set *sub-*$\mathscr{g}$ + $C$ $\pm$ $\Delta C$, and then the cycle $\Delta C$ is a missing cycle either or a redundant cycle. Note $G$ is a norm solvable graph. So we have $sub\text{-}g + C \pm \Delta C = S_x$ where $| C | \neq 0$ and $| \Delta C | \neq 0$. ■

## 3.3 The $\beta_2$-construction

### 3.3.1 $\beta$ and $\beta_2$ on $\Delta C$

According to Lemma 2, if comparing with the order and number of cycles in $S_x$, we can then find out the $\Delta C$. Thus, the next step is to identify the selected cycle as a $\Delta C$ to satisfy $sub\text{-}g+C\pm\Delta C = S_x$. The method to identify $\Delta C$ we proposed is the $\beta_2$-construction, in which the vertices $\beta$ on the selected cycle can be reduced to $\beta_2$ if we can delete all the basal cycles $C_\Delta$ on $\beta$ except $\Delta C$.

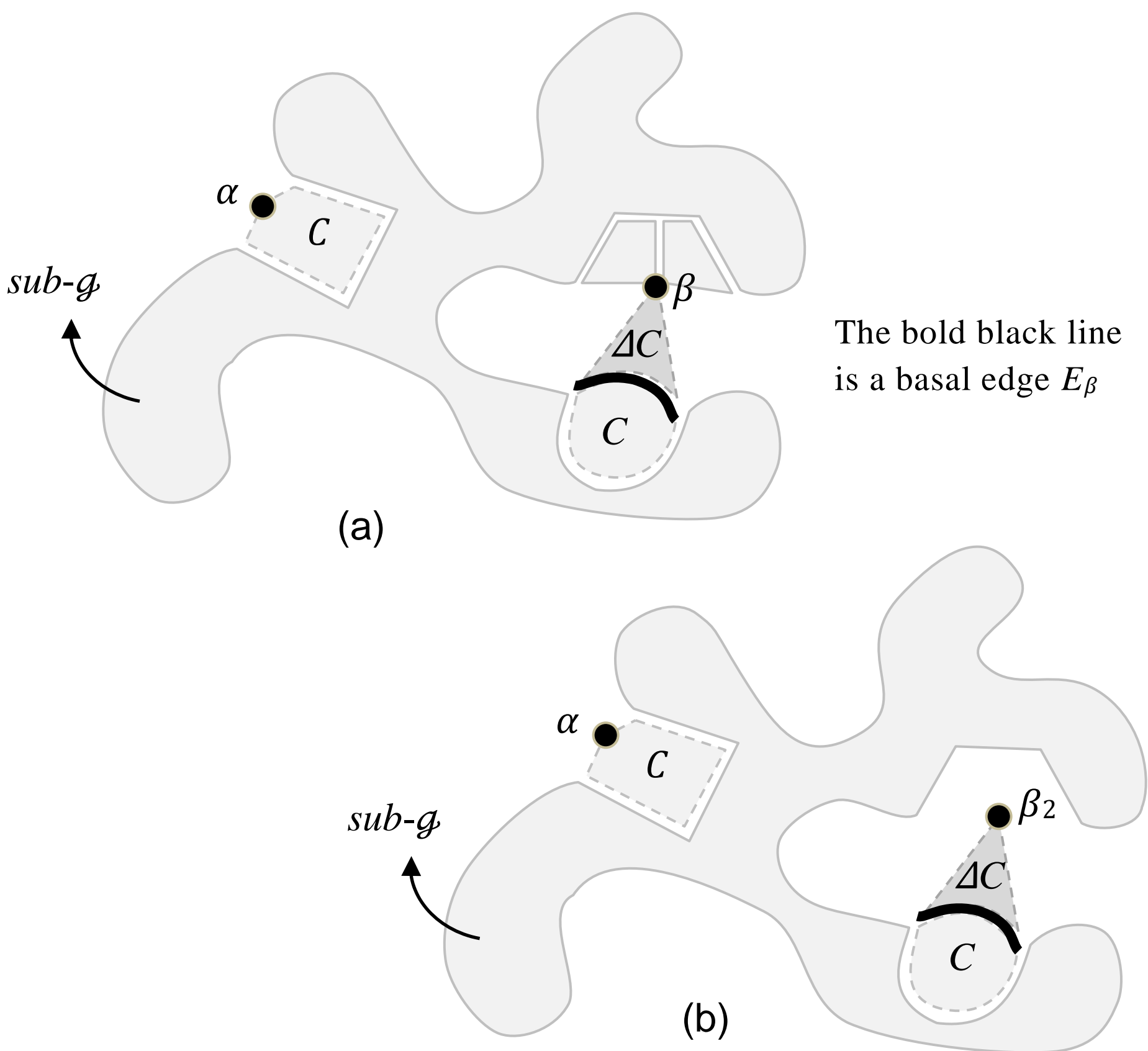

Figure 2. A diagram of $\beta_2$-construction in the set $sub\text{-}g$ (a) We select a cycle (dark grey) to be as $\Delta C$ connected with cycle $C$. (b) To identify that the selected cycle is the cycle $\Delta C$, by showing that there is a vertex $\beta_2$ (that is $|\beta_2|\neq 0$).

**Proposition 3** *If $\beta = \bowtie$ in a $\beta_2$-construction, then $|\beta_2| = 0$.*

*Proof.* By the definition, for a vertex $\beta$ on the selected cycle, if it is a cut vertex, then, with $\beta$ being a common vertex, any connected cycle cannot be $C_\Delta$. By the definition, the connected cycle must be a cycle having a vertex of degree 2. Clearly, the vertex $\beta$ cannot be reduced to a vertex $\beta_2$, thus $|\beta_2| = 0$. ■

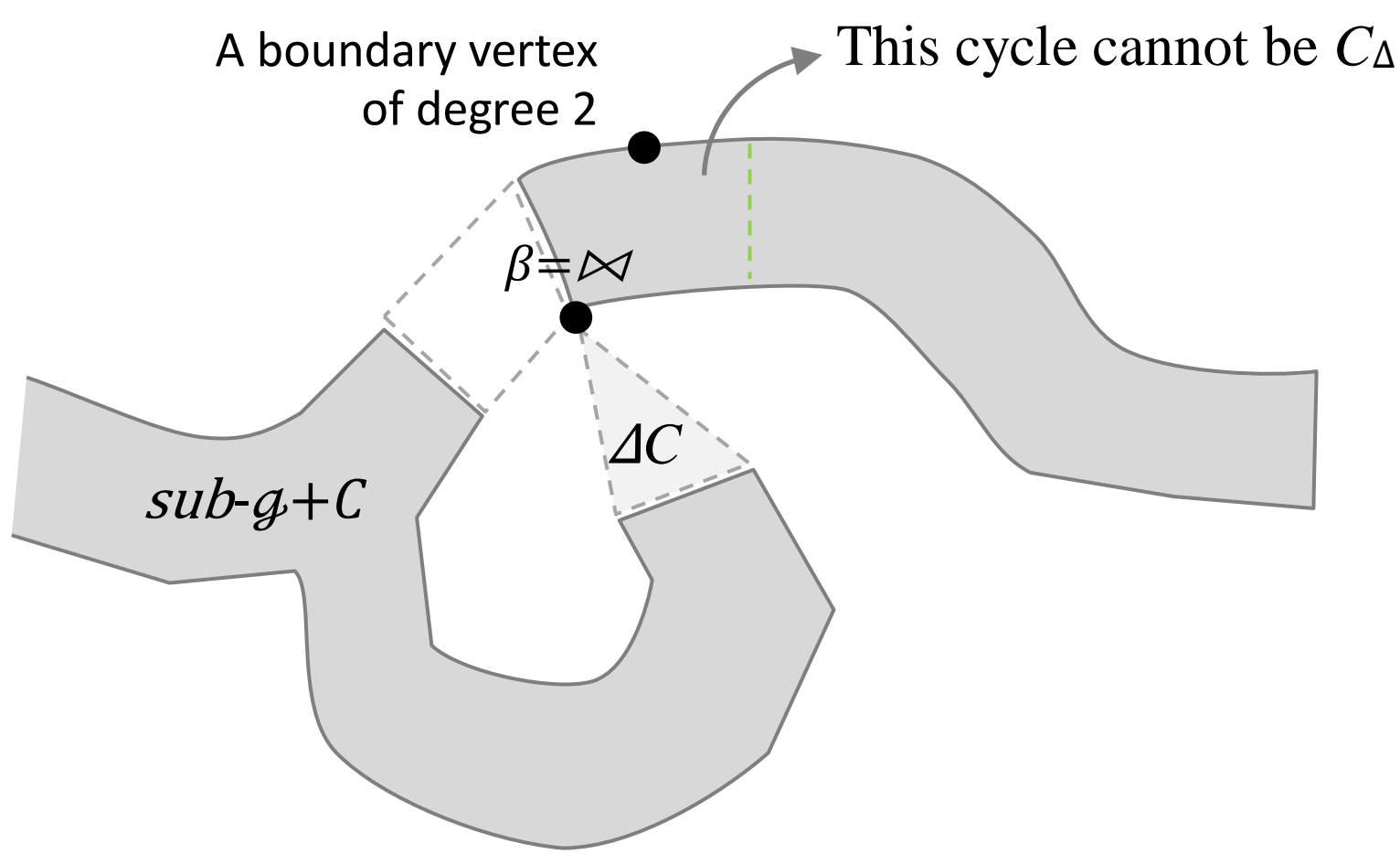

Figure 3. $\beta=\bowtie$ in the $\beta_2$-construction on $\Delta C$

For $sub\text{-}\mathscr{g}+C \pm \Delta C = S_x$ where $|C| \neq 0$, if $|\beta_2| = 0$, we have

**Lemma 4** *Let G be a norm solvable graph. $S_x$ is a solution of G such that $S_x=\{f_i|\ 3\leq i\leq |V(G)|$ and $\sum(i|f_i|-2|f_i|)=|V(G|)-2\}$.The $sub\text{-}\mathscr{g}+C$ is a set of G such that $sub\text{-}\mathscr{g}+C = S_x$. Let $\Delta C$ be an unassigned solution cycle in an adding process. Then, G is non-Hamiltonian, if $|\beta_2| = 0$ for $sub\text{-}\mathscr{g}+C \pm \Delta C = S_x$ where $|C| \neq 0$.*

*Proof.* Consider $sub\text{-}\mathscr{g}+C \pm \Delta C = S_x$ where $|C| \neq 0$. By the definition, for a selected cycle, $|\beta_2| = 0$ gives two cases: there is no cycle can be selected to be as $\Delta C$, then $|\Delta C| = 0$; or there has $\beta=\bowtie$ on the selected cycle, by Proposition 3, $|\Delta C|=0$. By Corollary 1, $G$ is not a Hamilton graph. That completes the proof. ■

For $sub\text{-}g+C \pm \Delta C = S_x$ where $|C| \neq 0$, if $|\beta_2| \neq 0$, we have the results in the following three sections.

### 3.3.2 $\Delta C$ and $p$-$\mathcal{C}$

According to [7], a solution of $G$ is a set of solution cycles, which is a 2-common ($v$, $e$) cycle set, such that the sum of these cycles is a Hamilton cycle. Clearly, this solution is a particular solution and all the solution cycles are $p$-$\mathcal{C}$. We use $g$-$C_i$ for a general solution cycle of order i, and $p$-$C_i$ a particular solution cycle of order $i$, where $3 \leq i \leq V(G)$.

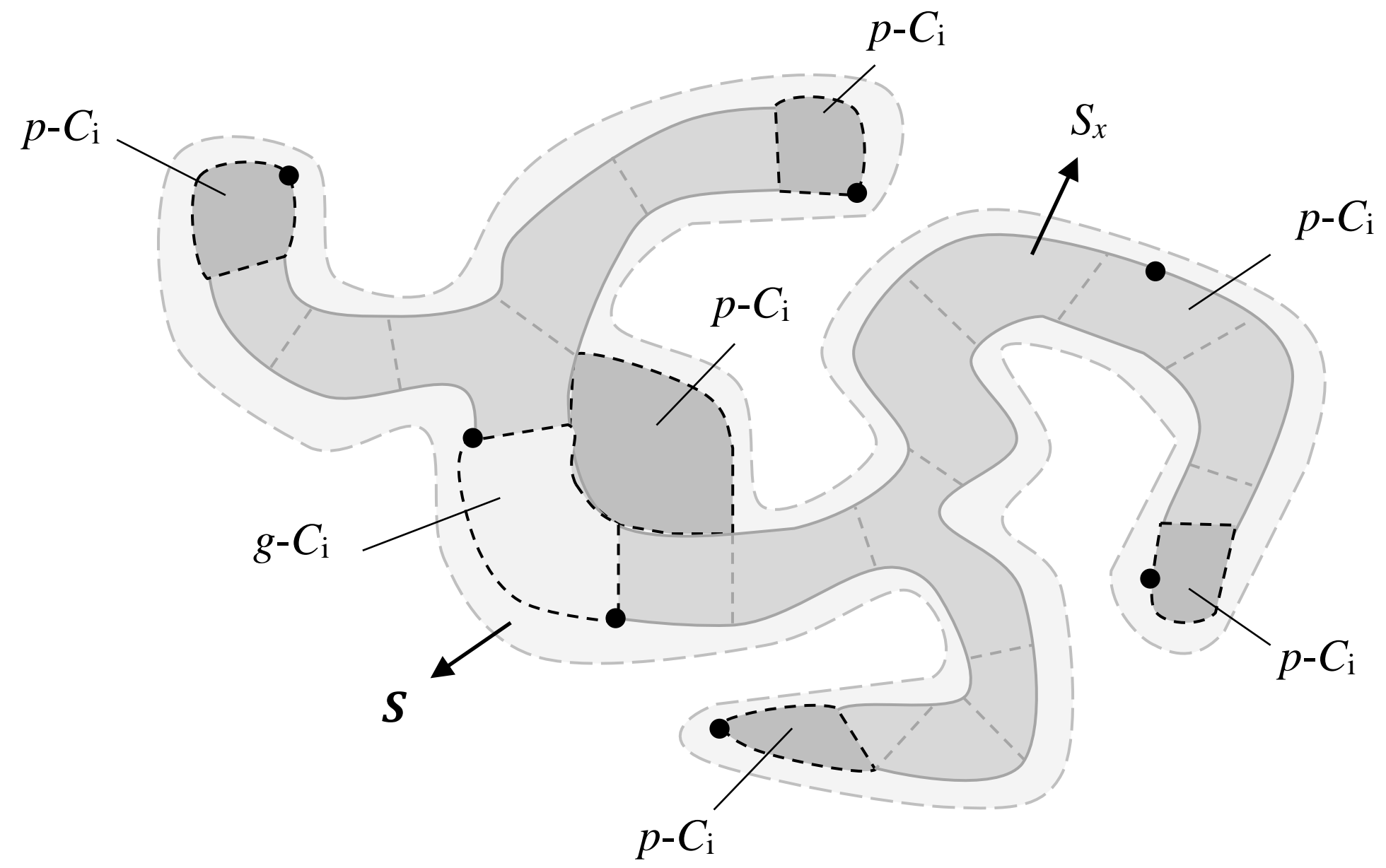

Figure 4. General solution cycles $g$-$C_i$ and particular solution cycles $p$-$C_i$

**Proposition 4** *Let g-$C_i$ be a general solution cycle of order i, and p-$C_i$ a particular solution cycle of order i, where $3 \leq i \leq V(G)$. Then, $\Delta C = p\text{-}C_i$, if $|\beta_2| \neq 0$ for all $\beta$ on a $\Delta C$.*

*Proof.* In a $sub\text{-}g+C \pm \Delta C$, $|\beta_2| \neq 0$ for all $\beta$ on a $\Delta C$ implies that $\Delta C$ is a solution cycle in a 2-common ($v$, $e$) cycle set or a 2-common ($v$, $0$) cycle set. Since all vertices $\beta$ can be reduced to vertices $\beta_2$, then, clearly, $\Delta C$ must be a $p$-$C_i$. Hence, we have $\Delta C = p\text{-}C_i$. ∎

### 3.3.3 The upper bound of searching all $\Delta C$

For finding suitable $\Delta C$, we need an ergodic searching for all the potential cycles. There is an upper bound of the searching algorithm.

**Lemma 5** *Let G be a norm solvable graph. Let* $\beta_2$ *denote a boundary* $\beta$ *of degree 2.* $|\beta_2| \neq 0$ *denotes that the vertex* $\beta$ *cannot be reduced to* $\beta_2$*. There have at most* $|V(G)|-2$ *cycles* $\Delta C$ *in a sub-𝓰+C.*

*Proof.* Suppose a *sub-𝓰+C* has a set of $\Delta C$ such that $V(\Delta C) \leq V(G)$. Since there is no inside vertex in the set of $\Delta C$, then it is clearly that there are at most $|V(G)|-2$ cycles $\Delta C$ in a *sub-𝓰+C*. ■

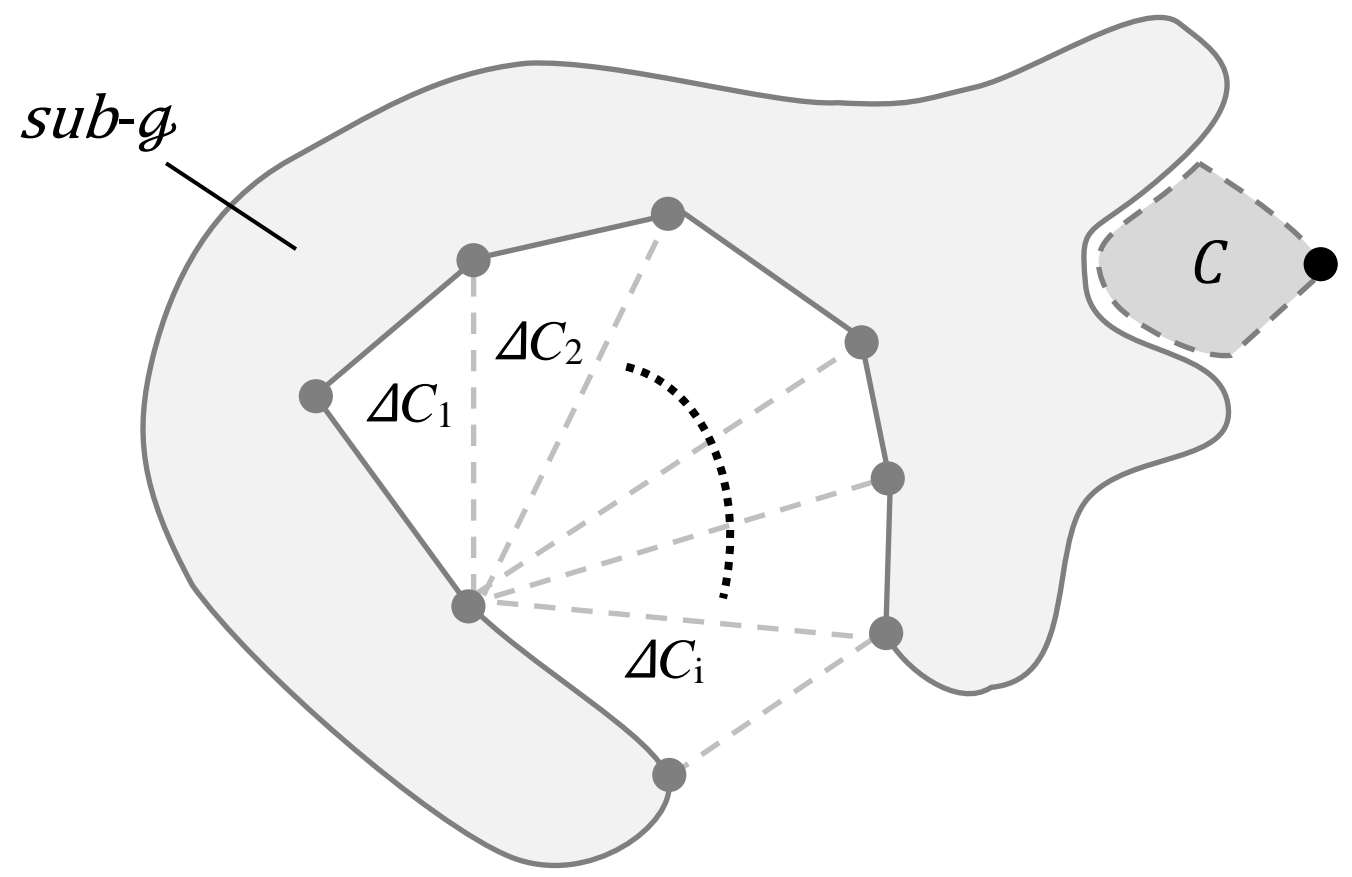

Figure 5. A diagram of the set of $\Delta C$ in the set *sub-𝓰*

### 3.4 The existence of $\alpha$

**Lemma 6** *Let G be a norm solvable graph. Let sub-𝓰 + C be a subset of G* such that *sub-𝓰* + $C \neq S_x$ *where* $|C| \neq 0$*. Then, G is a non-Hamilton graph, if* $|\alpha| \neq 0$ *and* $|\beta_2| \neq 0$ *for all* $\Delta C$ *in the* $\beta_2$*-construction of the sub-𝓰* + $C \pm \Delta C$*.*

*Proof.* Since every $\beta_2$-construction on one $\beta$ is independently, then, without loss the generality, we consider the case on one $\beta$.

If $|\alpha| \neq 0$ and $|\beta_2| \neq 0$ in the $\beta_2$-construction on a vertex $\beta$ for any $\Delta C$, then there must emerge an isolated vertex $\alpha$, and we have

$V(sub\text{-}g+C\pm\Delta C-C_\Delta) < V(G)$, and $sub\text{-}g+C\pm\Delta C-C_\Delta \neq S_x$. Then, $\Delta C \neq p\text{-}C_i$ for the set $sub\text{-}g+C\pm\Delta C$ and $\Delta C$ cannot satisfy the match. The result means that the selected cycle cannot be a $\Delta C$, that is $\Delta C$=0. Hence, for all $\Delta C$ in the $\beta_2$-construction, every case of $|\alpha|\neq 0$ and $|\beta_2|\neq 0$ implies that there is no cycle set with $\Delta C$ in $G$ to match $S_x$. Thus, $G$ is a non-Hamilton graph. ■

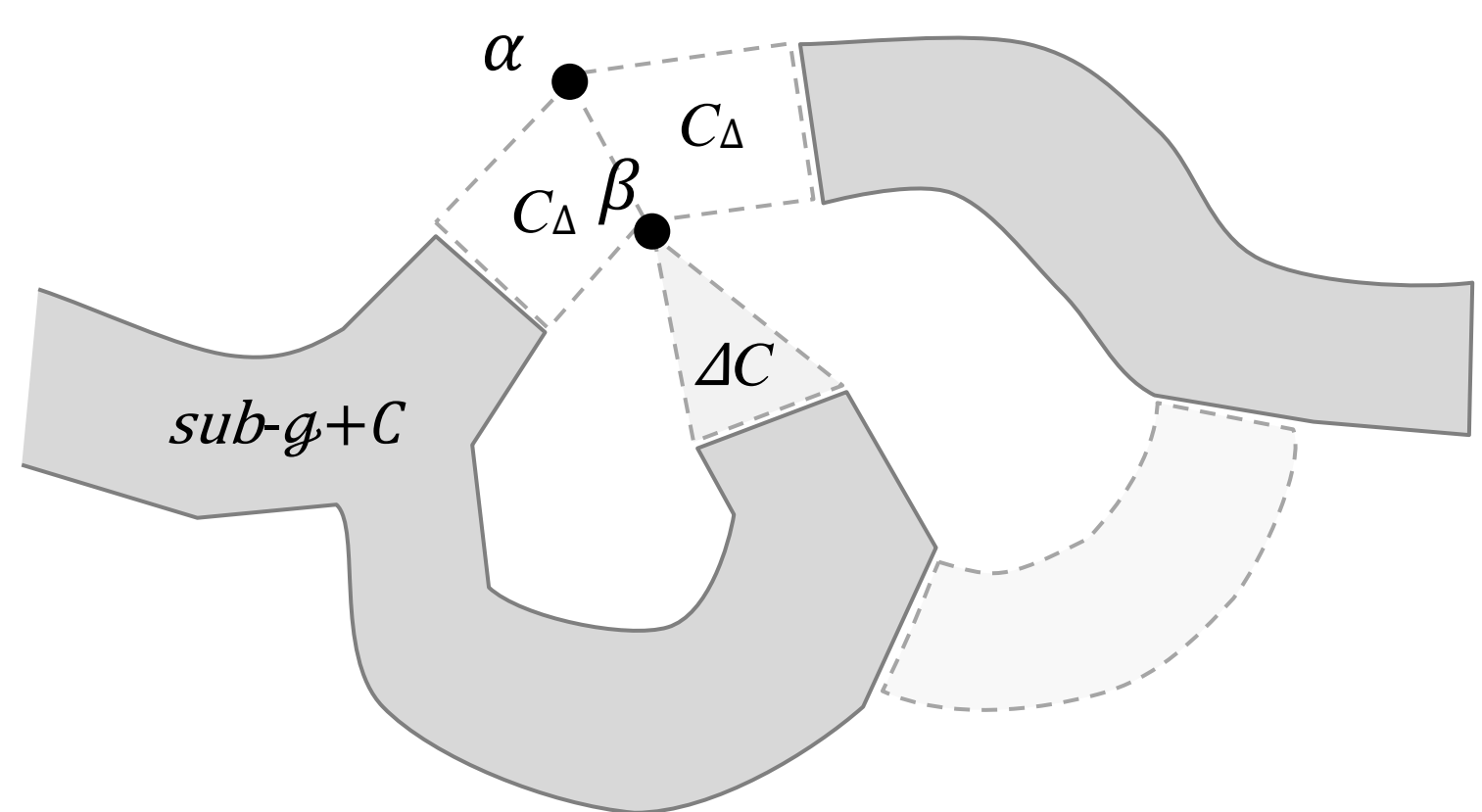

Figure 6. An isolated vertex $\alpha$ generated from of the $\beta_2$-construction on $\Delta C$ in the set $sub\text{-}g+C$

**Lemma 7** *Let G be a norm solvable graph. Let* $sub\text{-}g+C$ *be a subset of G* such that $sub\text{-}g + C \neq S_x$ *where* $|C| \neq 0$. *Then, G is a Hamilton graph, if* $|\alpha| = 0$ *for all* $\Delta C$ *and* $|\bowtie| = 0$ *on* $C_\Delta$ *in the* $\beta_2$*-construction of the* $sub\text{-}g+C \pm \Delta C$, *where* $\beta \neq \bowtie$.

*Proof.* For any $\beta$ in the $\beta_2$-construction, $|\alpha|=0$, $\beta\neq\bowtie$, $|\beta_2|\neq 0$, and $|\bowtie| = 0$ on $C_\Delta$ means that deleting all $C_\Delta$ does not leave any vertex as an isolated vertex $\alpha$. We only need to consider two sub-cases: $|\omega|=1$ and $|\omega|=2$.

Sub-case $|\omega|=1$, $|\alpha|=0$, $\beta\neq\bowtie$, and $|\beta_2|\neq 0$. There has only one structure in this sub-case. Clearly, when cycle $C$ being deleted as a $C_\Delta$, we obtain a 2-common $(v, e)$ cycle set. Thus, for all $\Delta C$ with the same sub-case independently, we derive a match with a 2-common $(v, e)$ cycle set of $G$. Thus, $G$ is a Hamilton graph.

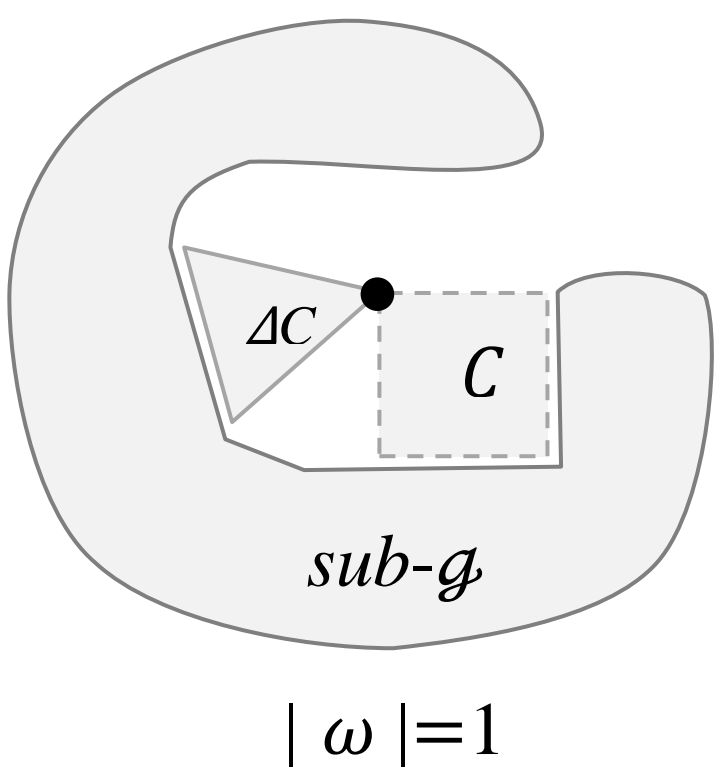

Figure 7. Sub-cases $|\omega|$=1 with $|\alpha|$=0, $\beta \neq \bowtie$, and $|\beta_2| \neq 0$ in the $\beta_2$-construction on $\Delta C$ in the set *sub-g*+$C$

Sub-case $|\omega|$=2, $|\alpha|$=0, $\beta \neq \bowtie$, and $|\beta_2| \neq 0$. We only consider $|\bowtie| = 0$ and $|\bowtie| \neq 0$ on $C_\Delta$.

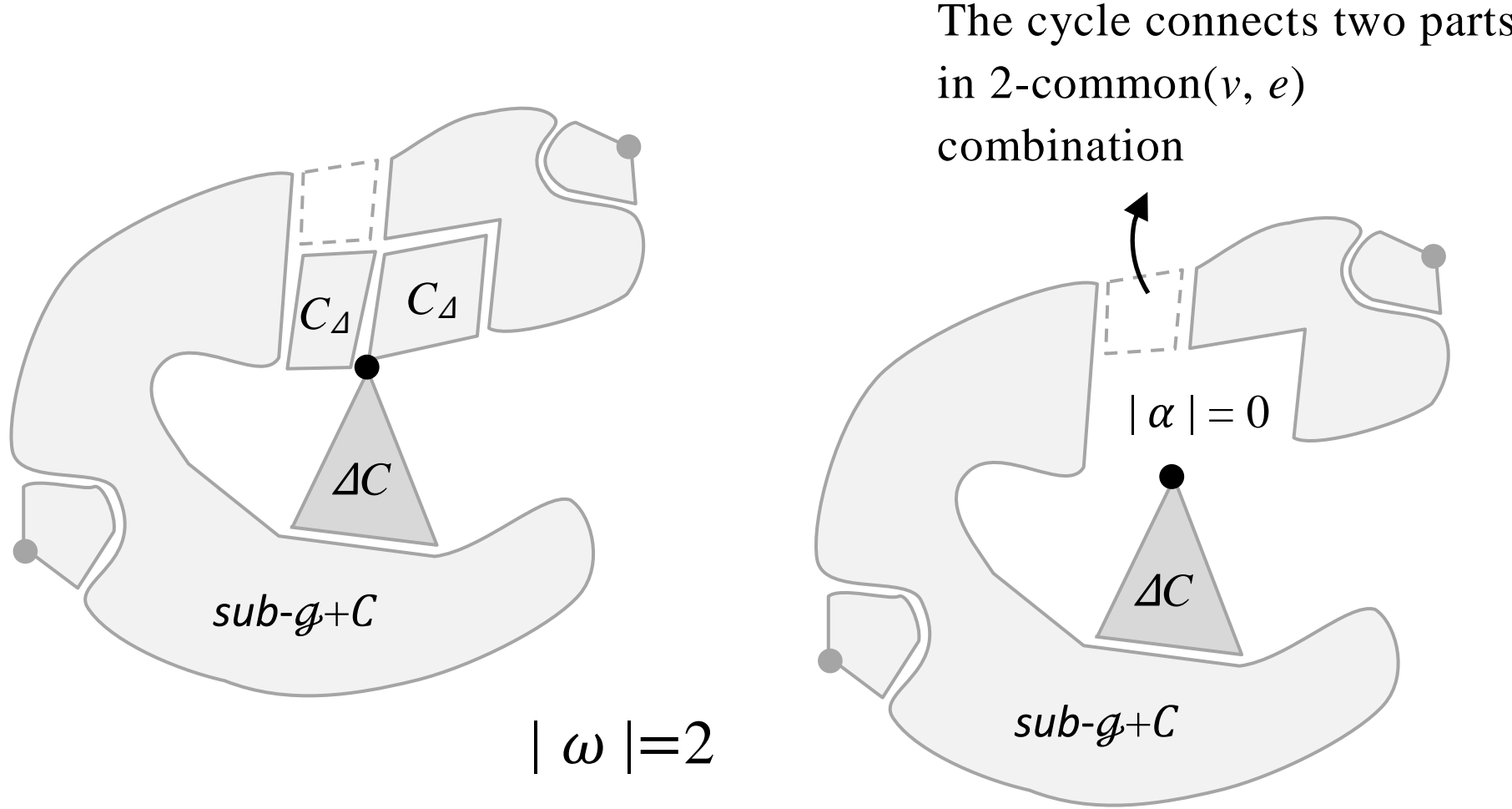

Figure 8. Sub-cases $|\omega|$=2 with $|\bowtie|$=0, $|\alpha|$=0, $\beta \neq \bowtie$, and $|\beta_2| \neq 0$ in the $\beta_2$-construction on $\Delta C$ in the set *sub-g*+$C$

For $|\bowtie|$= 0, since $G$ is a solvable graph and $|\alpha|$=0, then, by the definition, there must exists a cycle, which not be selected for producing the *sub-g*, connecting two parts in the way of the 2-common ($v$, $e$) combination. This implies that the connection produces a new set *sub-g*+$C$+$\Delta C_i$ where $i = t$ and $1 \leq t \leq |B(G)|-|S_x|$. Clearly, $\Delta C_i$ is a particular solution cycle, $\Delta C_i = p\text{-}C_i$. Note that $|\alpha|$=0 for all $\Delta C_i$, so we have *sub-g* + $C$ + $\Delta C_i = S_x$ and the *sub-g* + $C$ + $\Delta C_i$ is a 2-common ($v$, $e$) cycle set of $G$. Thus, $G$ is a Hamilton graph.

We complete the proof. ■

**Lemma 8** *Let G be a norm solvable graph. Let* sub-$g$+$C$ *be a subset of G* such that sub-$g$+$C \neq S_x$ *where* $|C| \neq 0$. *Then, G is non-Hamilton, if* $|\alpha|=0$ *for all* $\Delta C$ *and* $|\bowtie| \neq 0$ *on* $C_\Delta$ *in the* $\beta_2$*-construction of the* sub-$g$+$C \pm \Delta C$, *where* $\beta \neq \bowtie$.

*Proof.* For $|\bowtie| \neq 0$ on $C_\Delta$, it implies that there two parts of the sub-$g$+$C$+$\Delta C$ after implementing the $\beta_2$-construction, one part includes $\Delta C$, the other part includes a cut vertex $\bowtie$. Note that $G$ is a norm solvable graph and the connecting cycle between two parts, by Lemma 3.1 in [8], there must exists a 2-common ($v$, $0$) cycle set of $G$, and by the definition in [7], the vertex $\bowtie$ is a vertex $K$. Hence, $G$ is non-Hamilton. ■

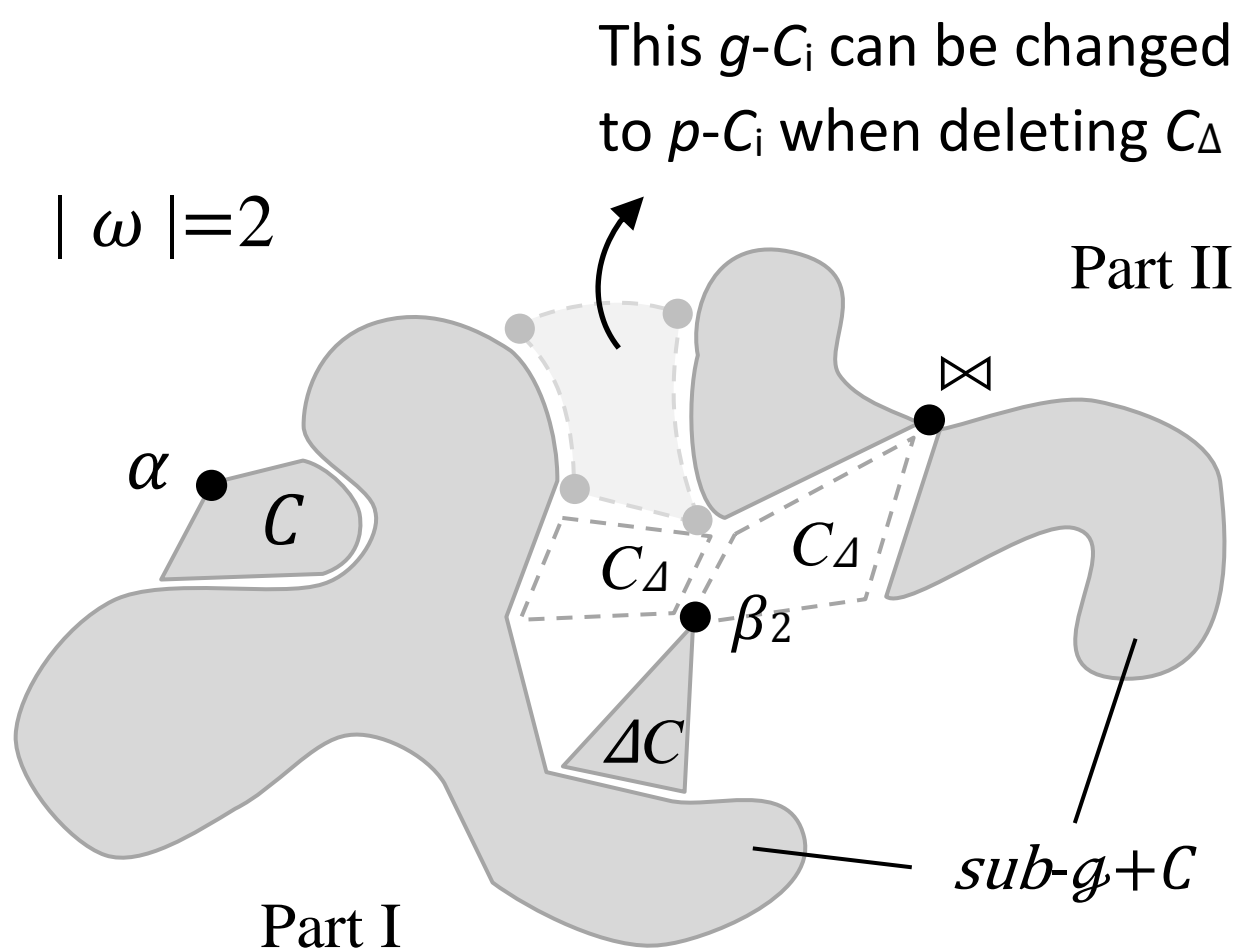

Figure 9. The sub-cases $|\omega|=2$ with $|\bowtie| \neq 0$ $|\alpha|=0$, $\beta \neq \bowtie$, and $|\beta_2| \neq 0$ in the $\beta_2$-construction on $\Delta C$ in the set sub-$g$+$C$

## 4 The "1+1" algorithm theorem and its proof

For most of algorithms to the Hamilton cycle problem, the general process is to input the data of sets of the vertices and edges of a graph, and then to obtain a Hamilton cycle. Accordingly, to implement the "1+1" algorithm, we need to input the data including the vertices and edges of a graph and transform the graph into a normal graph; and then we obtain the minimal cycle base from the normal graph, and solve the equation of the given graph.

Finding the minimal cycle base and the solutions has polynomial algorithms [9, 10]. And, it is easy to estimate the polynomial complexity of the algorithm for normalization of $G$. Note that these processes are independent respectively. Hence, these processes overall cost polynomial time. Therefore, in this paper, we simplify the designation of our algorithm to concentrate on the difficult parts in HCP algorithm. Thus, we suppose $B(G)$, $\boldsymbol{S}$, and $S_x$ are given for the algorithm and give the theorem of the "1+1" algorithm as follow.

**Theorem 1.** *Let G denote a solvable normal graph. Let B(G) be the minimal cycle base of G.* $\boldsymbol{\mathcal{S}}$ *is a general solution, and* $S_x$ *a solution. The "1+1" algorithm is an* $\mathcal{O}(|E(G)|^3)$ *worst-case time algorithm for finding a Hamilton cycle.*

*Proof.* Our proof has two parts.

Part 1 The performing the "1+1" algorithm yields $G$ is a Hamilton graph, either or, $G$ is a non-Hamilton graph.

By **Corollary 1.1** [6], any graph without solutions is non-Hamiltonian. So, in our algorithm we consider solvable normal graphs only.

The aim of the "1+1" algorithm is going to construct a cycle set comparing to $\mathcal{g}$, so that we can determine whether the given graph is Hamiltonian or not by **Theorem 3.3**[4] [8]. By the definition of $\mathcal{g}$ and $G$ is a solvable norm graph, we have $\mathcal{g}=S_x$. So, our approach to constructing $\mathcal{g}$ is to produce a solution set $S_x$.

In constructing, we first obtain a 2-common $(v, e)$ cycle set, a sub-$\mathcal{g}$, by adding every sequential cycle in the way of 2-common $(v, e)$ combination. And then, by utilizing the methods of "adding $C$",

[4] **Theorem 3.3** G is Hamiltonian $\Leftrightarrow \mathcal{g} \approx K_{2,3}$.

"adding $\pm\Delta C$", and "$\beta_2$-construction", our algorithm gives an output that $G$ is a Hamilton graph or a non-Hamilton graph.

The steps of the "1+1" algorithm are given as follows.

---

**Algorithm:** the "1+1" algorithm for HCP

---

**Input:** $S_x = \{C_1, C_2, \ldots, C_t\}$, where $x \in \mathbb{Z}^+$, $1 \leq t \leq |B(G)|$.
**Output:** $G$ is either a Hamilton graph or a non-Hamilton graph.

S1: Initialize the arguments

count_*sub-𝒼* = 0

count_$V$_ *sub-𝒼* = 0

count_$C$ = 0

count_$C_k$ = 0

count_*sub-𝒼* +$C$ = 0

count_$\beta_2$ = 0

count_$\alpha$ = 0

count_$V$_ *sub-𝒼* +$C\pm\Delta C$ = 0

count_$\bowtie$ = 0

S2: Enumerate a cycle from $\{C_1, C_2, \ldots, C_t\}$ to produce the *sub-𝒼* by the definition, until no satisfied cycle in $S_x$

S3: Detect $C$ for isolated vertex $\alpha$. By **Proposition 2**, the existence of $C$ can be used to express the existence of isolated vertex $\alpha$. Comparing the orders of the *sub-𝒼* and $S_x$, we can determine the existence of $C$.

We begin from computing count_$S_x$ and count_*sub-𝒼* to sift the one-cycle graphs.

**if** count_ $S_x$ = 1 and count_*sub-𝒼* = 1, **then**

by the definition, we have count_$C$ = 0

by **Proposition 1**, print "$G$ is a Hamilton graph"

**else**

count_ $S_x$ > 1 and count_*sub-𝒼* > 1, **then**

compute the number of R =1 edges to represent the order of *sub-𝒼*, and compare to that of $S_x$

**if** count_$V$_ *sub-𝒼* = count_$V$_ $g$-$\boldsymbol{C}$, **then**

we have count_$C = 0$
by **Proposition 1**, print "*G* is a Hamilton graph"
**else**
goto S4
**end if**
**end if**

S4: Check the combination of every $C$ and *sub-g* to identify the existence of $C_k$ by the definition
**if** count_$C_k = 0$, **then**
goto S5
**else**
count_$C_k \neq 0$
by **Lemma 3.3**[5] [7], print "*G* is non-Hamiltonian"
**end if**

S5: Match the *sub-g* + $C$ to $S_x$. Compute the number of cycles in the *sub-g* + $C$ and $S_x$, expressed by count_*sub-g*+$C$ and count_$S_x$ respectively
**if** count_*sub-g* +$C$ = count_$S_x$, **then**
count_$\varDelta C = 0$, which means there is a match for two sets
*sub-g*+$C$ =$S_x$
by **Corollary 1**, print "*G* is non-Hamiltonian"
**else**
count_*sub-g* +$C$ = count_$S_x$, and
count_$\varDelta C \neq 0$, which means there is no match
goto S6
**end if**

S6: Test the basal cycle $C_i$ on every $\varDelta C$ to find $C_k$
**for** every $\varDelta C$ **do**
check vertex $k$ on $C_i$ when adding a $\varDelta C$
**if** $C_i = C_k$, **then**
count_$C_k \neq 0$
by **Lemma 3.3** [7]
print "*G* is non-Hamiltonian"

[5] **Lemma 3.3** *G* is Hamiltonian ⇔|$C_k$|=0.

**else**

$C_i \neq C_k$

count_$C_k$ = 0, goto S7

**end if**

**end for** every $\Delta C$

S7: Perform the $\beta_2$-construction on every $\Delta C$

Delete all the cycles without relations to a basal cycle $C_\mathrm{i}$ on $\Delta C$ and observe all generated vertices

**for** every $C_\mathrm{i}$ on $\Delta C$ **do**

delete all the cycles without relations to a basal cycle $C_i$

**if** all the vertices are degree 2, **then**

count_$\beta_2 \neq 0$, goto S8

**else**

there is a vertex of degree 2

count_$\beta_2$ =0

By **Lemma 4**, print "$G$ is non-Hamiltonian"

**end if**

**end for** every $C_i$ on $\Delta C$

S8: **do**

compute the number of $R$ = 1 edges in the $\mathit{sub}\text{-}g+C\pm\Delta C$ and $S_x$, and test the existence of isolated vertices after producing $\beta_2$, which are referred by count_$R1$_$\mathit{sub}\text{-}g+C\pm\Delta C$ and count_$R1$_$S_x$ respectively

count_ $R1$_$\mathit{sub}\text{-}g+C\pm\Delta C \neq$ count_$R1$_$S_x$

there has an $\alpha$, and

count_$\alpha \neq 0$

by **Lemma 6**, print "$G$ is non-Hamiltonian"

**while** count_ $R1$_$\mathit{sub}\text{-}g+C\pm\Delta C$ = count_$R1$_$S_x$,

there have no $\alpha$

count_$\alpha$ =0

goto S9

**end do**

S9: Search to detect the cuts $\bowtie$ on all $C_\Delta$

**do**

compute the number of vertices in the $sub\text{-}\mathscr{g}+C\pm\Delta C$ and $S_x$, which are represented by count_V_$sub\text{-}\mathscr{g}+C\pm\Delta C$ and count_V_$S_x$. By **Lemma 1**, we have

count_$V$_ $sub\text{-}\mathscr{g}+C\pm\Delta C$ = count_$V$_$S_x$, no $\bowtie$

count_$\bowtie$ =0

by **Lemma 7**, print "$G$ is Hamiltonian"

**while** count_$V$_$sub\text{-}\mathscr{g}+C\pm\Delta C \neq$ count_$V$_$S_x$, there is a $\bowtie$

count_$\bowtie \neq 0$

by **Lemma 8**, print "$G$ is non-Hamiltonian"

**end do**

---

From S1 to S2, we obtain a $sub\text{-}\mathscr{g}$ from the given graph $G$, a solvable normal graph. For the purpose of comparison, we will implement several steps to prone the $sub\text{-}\mathscr{g}$ to a $sub\text{-}\mathscr{g}+C$ and a $sub\text{-}\mathscr{g}+C+\Delta C$. By **Lemma 2** and **Lemma 3**, we can match the $sub\text{-}\mathscr{g}+C$ or the $sub\text{-}\mathscr{g}+C+\Delta C$ to $S_x$, and then to $\mathscr{g}$. However, in the procedure of constructions and comparisons, the program continuously sifts Hamilton and non-Hamilton graphs out from the working sets until all of them are fixed.

In S3, by the iteration of searching edges for every cycle in $sub\text{-}\mathscr{g}$, if no isolated vertex $\alpha$ left in generating the $sub\text{-}\mathscr{g}$, that is $|C|$=0, then, by **Proposition 1**, the program gives that the $sub\text{-}\mathscr{g}$ is Hamiltonian and then $G$ is Hamiltonian. Clearly, by **Proposition 2**, $C$ *is a g-*$\boldsymbol{\mathcal{C}}$. Therefore, the working set for the comparison remains

$$sub\text{-}\mathscr{g}+C \text{ where } |C|\neq 0. \quad (1)$$

In S4, the program will investigate if there is a $C_k$ generated when adding $C$ into the $sub\text{-}\mathscr{g}$. By **Lemma 3.3** in [7], if $|C_k|\neq 0$, then the program outputs $G$ is non-Hamiltonian. Then, the working set reduces to

$$sub\text{-}\mathscr{g}+C \text{ with } |C_k|=0, \quad (2)$$

and the program goes to S5.

S5 is to test the working set (2) to be a match for a solution set $S_x$.

By the iteration of selecting every cycle in $S_x$ to compare with the cycles in the working set (2), we can certain the existence of $\Delta C$. By **Corollary 1**, from $|\Delta C|=0$, we derive an output $G$ is non-Hamiltonian. Then the working set (2) converts to the following,

$$sub\text{-}\mathcal{g}+C\pm\Delta C \text{ with } |C|\neq 0,\ |C_k|=0 \text{ and } |\Delta C|\neq 0. \tag{3}$$

The unassigned cycle $\Delta C$ in the working set (3) will be added or deleted by the following steps 6, 7 and 8, so that we can determine the existence of $\beta_2$ and $\alpha$.

In S6, the program begins from the case of the adding cycle $\Delta C$. While selecting a $C_i$ in the set $S_x-(sub\text{-}\mathcal{g}+C)$ as an adding cycle $\Delta C$, we need first to investigate again the existence of cycle $C_k$. Clearly, by **Lemma 3.3** in [7], for $|C_k|\neq 0$, the program plainly outputs $G$ is non-Hamiltonian. Therefore, we derive a new working set

$$sub\text{-}\mathcal{g}+C\pm\Delta C \text{ where } |C_k|=0 \text{ for } +\Delta C,\ |C_k|=0 \text{ for } +C. \tag{4}$$

Then the program goes on.

In the next steps, we perform the $\beta_2$-construction on all the selected $C_i$ to examine the existence of isolated vertices $\alpha$ when reducing $\beta$ to $\beta_2$.

S7 is to identify the existence of $\beta_2$. If $|\beta_2|=0$, by **Lemma 4**, the program outputs $G$ is non-Hamiltonian. We then have the working set

$$sub\text{-}\mathcal{g}+C\pm\Delta C \text{ with } |\beta_2|\neq 0, \tag{5}$$

and go to the next.

In S8, we determine the existence of the isolated vertices $\alpha$ on every $\Delta C$ with $|\beta_2|\neq 0$. By **Lemma 6**, if $|\alpha|\neq 0$ for all $\Delta C$, then the program gives $G$ is non-Hamiltonian. We then obtain the working set

$$sub\text{-}\mathcal{g}+C\pm\Delta C \text{ with } |\alpha|=0 \text{ and } |\beta_2|\neq 0. \tag{6}$$

The working set (6) remains the last situation we need to consider that is the number of components in $\beta_2$-construction. We go to S9.

In S9, for the working set (6) of $|\omega|=1$, if $|\bowtie|=0$ on $C_\Delta$, by **Lemma 7**, the program outputs $G$ is Hamiltonian. Therefore, we derive the left working set

$$sub\text{-}\mathcal{G}+C\pm\Delta C \text{ with} |\alpha|=0,\ |\beta_2|\neq 0 \text{ and } |\bowtie|\neq 0 \text{ for } |\omega|>1. \quad (7)$$

By **Lemma 8**, the program outputs $G$ is non-Hamiltonian from the working set (7). Up to now, we have finished testing all the possibilities of the working set and gave the decision to the Hamiltoncity of $G$. The procedure ceases.

Part 2 The time complexity of the "1+1" algorithm is $\mathcal{O}(|E(G)|^3)$.

In S2, after arbitrarily selecting a cycle from $S_x$ as the beginning cycle, we need to enumerate a cycle from $t-1$ cycles in $S_x$ to construct a 2-common $(v, e)$ combination by comparing the all edges of the beginning cycle and the enumerated cycle. So the worst case of S2 is less or equal to $|E(G)|\cdot(t-1)$.

In S3, it is clearly that computing count_$S_x$ and count_$sub\text{-}\mathcal{G}$ costs $2t$ times. While to compute the number of $R$ =1 edges in $S_x$ and the $sub\text{-}\mathcal{G}$ we need $2|E(G)|$ times.

In S4, for detecting $C_k$ we need to find a cut on every $C$. With the sum of edge sets of $C$ and every cycle in the $sub\text{-}\mathcal{G}$, we could find $C_k$, which cost $(|E(G)|\cdot(t\text{-}1))\cdot|C|$ times.

In S5, to match the $sub\text{-}\mathcal{G}$ + $C$ and $S_x$, we need $2|S_x|$ times for matching cycles' number, and $|E(G)|\cdot|S_x|$ times for cycles' order. The total times is $2|S_x| + |E(G)|\cdot|S_x|$.

In S6, similar to the assessment in S4, The times cost includes to detect the $C_i$ is a cycle without boundary edges and to identify the $C_i$ is a co-solution cycle, which are $(|E(G)|\cdot(t\text{-}1))\cdot|\Delta C| + |E(G)|$.

In S7, for implements of the $\beta_2$-construction on every $\Delta C$, we first select a cycle $\Delta C$, and delete all the cycles without relations to a basal cycle $C_{\text{i}}$ on $\Delta C$, that need the times cost $|\Delta C|\cdot|E(G)|$. Secondly, we need to test whether the vertices, with no relations to the basal cycle $C_i$, are $\beta_2$ or not, that costs $|E(G)|\cdot(t-1)\cdot|\Delta C|$ times. So the total time cost is $|\Delta C|\cdot|E(G)| + (|E(G)|\cdot(t\text{-}1))\cdot|\Delta C|$.

In S8, we use every edge of one cycle to test each other cycle in the $sub\text{-}\mathscr{g}+C\pm\Delta C$ for finding the number of R=1 edges of the $sub\text{-}\mathscr{g}+C\pm\Delta C$, which costs $|E(G)|\cdot(t-1)$ times.

In S9, for finding the cuts $\bowtie$ on all $C_\Delta$, our approach is to compute the number of vertices in the $sub\text{-}\mathscr{g} + C \pm \Delta C$ and $S_x$, and this costs $|E(G)|\cdot|C_\Delta|$ times.

Since the procedure from S2 to S9 is a serial operation, hence, the overall time cost should be

$$|E(G)|\cdot(t-1) + 2|E(G)| + (|E(G)|\cdot(t\text{-}1))\cdot|C| + (2|S_x| + |E(G)|\cdot|S_x|)$$
$$+ [(|E(G)|\cdot(t\text{-}1))\cdot|\Delta C| + |E(G)|] + [|\Delta C|\cdot|E(G)| + |E(G)|\cdot(t-1)\cdot|\Delta C|]$$
$$+ |E(G)|\cdot(t-1) + |E(G)|\cdot|C_\Delta|.$$

Note that $t\le|B(G)|\le |E(G)-V(G)+1|$, $|C| \le |B(G)|+1\le|E(G)-V(G)+1|+1$, $|\Delta C| \le |V(G)|-2$, $|S_x| \le |B(G)| \le |E(G)-V(G)+1|$, and $|C_\Delta| \le |V(G)|-2$. Thus, the time complexity of the "1+1" algorithm is $\mathcal{O}(|E(G)|^3)$.

We complete the proof. ■

## 5 Conclusions

In this paper, based on the results in [6-8], we show some new Hamiltonian properties associated with the cycle base of solvable norm graphs, such as the relations between the general solution cycles and the particular solution cycles, the $sub\text{-}\mathscr{g}+C\pm\Delta C$ and $S_x$, and the $\beta_2$-construction and the isolated vertices $\alpha$. These properties provide serious methods for pruning working sets in different phases to match to a subgraph $\mathscr{g}$ (denoted by $S_x$ in this paper). Combining these methods, we derive the "1+1" algorithm for the Hamilton cycle problem and show it terminates in $\mathcal{O}(|E(G)|^3)$ time complexity.

## References

[1] Edmonds, J.: Minimum partition of a matroid into independent subsets. Journal of Research of the National Bureau of Standards 69B (1-2), 67-72 (1965)

[2] Gould, Ronald J.: Recent advances on the Hamiltonian problem: survey III. Graphs and Combinatorics (30), 1-46 (2014)

[3] Karp, R.M.: Complexity of Computer Computations. Springer, Boston, MA, Miller R.E., Thatcher J.W., Bohlinger J.D. (eds) 85-193 (1999)

[4] Basil Vandegriend: Finding Hamiltonian Cycles: Algorithms, Graphs and Performance. Master's thesis, University of Alberta (1998)

[5] Andrew Chalaturnyk: A Fast Algorithm for Finding Hamiltonian Cycles. Master's thesis, the University of Manitoba (2008)

[6] A note on the Grinberg condition in the cycle spaces https://arxiv.org/abs/1807.10187

[7] On non-Hamiltonian cycle sets of having solutions https://arxiv.org/abs/1906.09678

[8] The induced subgraph $K_{2,3}$ in a Non-Hamiltonian graph https://arxiv.org/abs/1906.10847

[9] J. D. Horton. A polynomial-time algorithm to find the shortest cycle basis of a graph. SIAM J. Comput. 16(2)

[10] L. G. Khchiyan, A polynomial algorithm in linear programming, Soviet Math Dokl., 20, 1979,191-194.
English version:
USSR Computational Mathematics and Mathematical Physics, 1980, 20:1, 53–72

[11] Reinhard Diestel, Graph Theory, Springer-Verlag, New York, Inc., 2000, 213.

[12] C. Berge, Graphs, Third revised ed., John Wiley & Sons, New York, 1991.

[13] J.A.Bondy, U.S.R.Murty, Graph Theory with Applications, Fifth Printing, Springer, 2008.

[14] Narsingh Deo, Graph Theory with Applications to Engineering and Computer Science. USA. Prentice-Hall, Inc., Englewood Cliffs, N. J., (1974).